\documentclass[prb,twocolumn,superscriptaddress,showpacs]{revtex4-1}
\usepackage{amssymb}
\usepackage{graphicx,epstopdf}
\usepackage{grffile}
\usepackage{tabularx}
\usepackage{amsmath}
\usepackage{amsfonts}
\usepackage{amssymb}
\usepackage{multirow}
\usepackage{pbox}

\begin{document}

\title{Ru-doping on iron based pnictides: the ``unfolded'' dominant role of structural effects for superconductivity}

\affiliation{Faculty of Physics, Center for Computational Materials Science, University of Vienna  (Austria)}
\affiliation{Department of Physical and Chemical Sciences, University of L'Aquila (Italy)}
\affiliation{Department of Physical and Chemical Sciences and SPIN-CNR, University of L'Aquila (Italy)}

\author {M. Reticcioli$^{1,2}$, G. Profeta$^3$, C. Franchini$^1$ and A. Continenza$^2$}

\begin{abstract}
We present an {\it ab-initio} study of Ru substitution in two different compounds, BaFe$_2$As$_2$ and LaFeAsO, pure and F-doped. Despite the many similarities among them, Ru substitution has very different effects on these compounds. By means of an unfolding technique, which allows us to trace back the electronic states into the primitive cell of the pure compounds, we are able to disentangle the effects brought by the local structural deformations and by the impurity potential to the states at the Fermi level. Our results are compared with available experiments and show: {\it i)} satisfying agreement of the calculated electronic properties with experiments, confirming the presence of a magnetic order on a short range scale; {\it ii)} Fermi surfaces strongly dependent on the internal structural parameters, more than on the impurity potential.
These results enter a widely discussed field in the literature and provide a better understanding of the role of Ru in iron pnictides: although isovalent to Fe, the Ru-Fe substitution leads to changes in the band structure at the Fermi level mainly related to local structural modifications.
\end{abstract}

\maketitle

\section*{Introduction}
\label{Intro}
Fe-based pnictides compounds constitute a large class of  materials, usually showing  superconductivity upon doping or applied pressure\cite{revSCstewart}.
As these  effects act on the electronic interactions, they have been both carefully studied and recognized to be powerful tools to gain insights into the properties of these materials.
In particular, doping has been extensively investigated since it allows tuning of the carrier number at the Fermi level; however, it also brings local distortions and disorder, introduces scattering centers and affects electronic correlations and magnetic properties. All these phenomena are of course entangled and often strictly dependent on the particular compound, so that it is very hard to single out how the different mechanisms cooperate to eventually drive the system into the superconducting state.
Earlier studies\cite{berlijn1, wadati, canfield} concentrating on doping, addressed the role of the additional charges brought about by the substitution and to its possible contribution to pair-breaking and disorder\cite{ba122unfold};
however, more recently\cite{Hajiri16,Fink15,Razzoli15,ba122bandw} the attention has been driven to the effects that different band-filling of selected states might have on electronic correlations and/or on magnetic fluctuations and to their possible consequences on the pairing mechanism. 
Moreover, the wide variety of Fermi surface (FS) topologies found in superconducting Fe-based materials\cite{ba122bandw} together with the presence/absence of band-nesting at the Fermi level, contributes to hinder a satisfying and unique picture that could explain the pairing mechanism in these materials.

In this context, the role of dopants is still very debated\cite{Merz} and  surely needs further studies. As an example, it is not clear why so-called {\it isovalent} substitutions are able to favor onset of superconductivity and how it is possible that the same dopant in different (but still similar) compounds  may induce completely different behaviors.
This is the case of Ru substituting on the transition metal (TM) sites of the Fe-As active layers in BaFe$_2$As$_2$ (Ba-122): Ru is considered to be isovalent with Fe, therefore no charge variations and almost unchanged Fermi surface features should be expected.
On the contrary, the electronic properties at the Fermi level are found to change considerably upon Ru-doping\cite{ARPES} bringing the compound into the superconducting state\cite{ba122Tc} which is then kept up to rather large Ru-concentrations (up to $\simeq$ 35\%), where disorder effects are expected to provide strong pair-breaking mechanisms.
On the other hand, Ru-doping in  LaFeAsO (La-1111) is seen not to induce superconductivity in the pure samples and to disrupt the superconducting state in the optimally F-doped (La-1111F) samples\cite{la1111Tc,daghero,la1111F-phase}, making the transition critical temperature ($T_C$) linearly drop with Ru-content.
At the same time, Ru-substitution in  pure La-1111  is seen to hinder magnetic order\cite{la1111struc1} while not inducing transition towards the superconducting state. Similar behavior was found for the Sm-1111 compound\cite{Iadecola,sanna,La1111Sm}.
Thus, the question arises: why Ru-doping has such different impact on 122 and 1111 systems, despite the many similarities between these compound families?

In the present work we concentrate on the effects of  structural changes induced by atomic substitution on the electronic properties of each compound and we will show that these need to be carefully investigated in order to find reasonable answers.
In the following, we will discuss the electronic properties of Ru-substitutions into Ba-122, La-1111 and La-1111F, performing {\it computational experiments} to single out pure electronic from pure structural contributions to the changes induced on the states at the Fermi level and comparing step by step the two cases considered, namely the Ba-122 and La-1111 compounds.

In order to analyze how the band structure of the pure compounds is affected by chemical substitutions, we make use of a computational technique  to unfold the supercell states into the larger primitive Brillouin zone, thus allowing to clearly compare the so called effective band structure (EBS) of the doped compound with the band structure of the pure compound and with available angle resolved photoemission spectroscopy (ARPES) experiments.

Our systematic study shows that:
{\it i)} states at the Fermi level in Ba-122 are very sensitive to structural changes due to Ru-substitution (namely, lattice parameters and the As-height with respect to the Fe-planes); 
{\it ii)} changes of these states result in orbital selective doping effects: hole-pockets are seen to be much more sensitive to structural changes compared to electron-pockets;
{\it iii)} the different spatial localization of Ru d-states, with respect to Fe, has little or very negligible effects on the Fermi topology, while its contribution becomes relevant only at higher binding energies on the density of states;
{\it iv)} Ru-substitution does not cause relevant changes to the local structure in La-1111; thus, its effect is much less evident and does not appreciably change the band structure at the Fermi level;
{\it v)} Ru d-states, nevertheless, affect TM-As hybridization, thus resulting into lower magnetic moments and magnetic order suppression, in both Ba-122 and La-1111 compounds.
 
The present paper is organized as follows: after a brief discussion of the computational approach in Sect.~\ref{compsetup}, we analyze the structural changes induced by Ru-substitution in Sect.~\ref{structure}, while in Sect.~\ref{electronic} we discuss the effects of structure and charge-doping on the states at the Fermi level for both compounds.
In Sect.~\ref{DOS} we discuss the differences between Fe and Ru on the global binding properties of the compounds and, finally, in Sect.~\ref{conclusions} we draw our conclusions.

\section{Computational details}
\label{compsetup}
We study Ru-substitutions on iron sites of BaFe$_2$As$_2$ (Ba-122), LaFeAsO (La-1111) and LaFeAsO$_{0.875}$F$_{0.125}$ (La-1111F) pnictides by means of an {\it ab-initio} approach within density functional theory (DFT).
The calculations are performed using the VASP\cite{kresse1,kresse2} package within the generalized gradient approximation\cite{Per96} (GGA) to density functional theory, and the projected augmented-wave (PAW)\cite{Blo94}.
Inclusion of 3\textit{p}-electrons in the valence shell of the iron atoms has been found to be substantially relevant for a more accurate description of the system\cite{colonna}.

In order to consider Ru-doping, we make use of the supercell approach. To this end, supercells 8 times larger than the primitive cells (including 2 Fe-atoms) have been considered; these supercells contain 40 and 64 atoms for Ba-122 and La-1111(F) respectively, and are defined as follows:
\begin{equation}
 \begin{cases}
\mathbf{A}_1 = 2 \mathbf{a}_2 + 2 \mathbf{a}_3 \\
\mathbf{A}_2 = 2 \mathbf{a}_1 + 2 \mathbf{a}_3 \\
\mathbf{A}_3 = \mathbf{a}_1 + \mathbf{a}_2
\end{cases}~\text{and}~
 \begin{cases}
\mathbf{A}_1 = 2 \mathbf{a}_1 + 2 \mathbf{a}_2 \\
\mathbf{A}_2 = -2 \mathbf{a}_1 + 2 \mathbf{a}_2 \\
\mathbf{A}_3 = \mathbf{a}_3
\end{cases}
\label{Avsa}
\end{equation}
where $\mathbf{a}_i$ and $\mathbf{A}_i$ are the lattice vectors of the primitive cell and the supercell respectively.
With this choice, each supercell includes 16 transition metal (TM) sites, i.e.\ 16 Fe/Ru atoms.

Different configurations of Ru-impurities have been considered in order to check their possible consequences on the electronic properties and on the states at the Fermi level. We found that changes in the Fe-site positions substituted by Ru atoms, at fixed Ru-concentration, do not affect appreciably the electronic states, as long as the Ru-occupation is limited to the metallic plane and equally distributed among different planes. The total energy of the system also is only slightly affected by the impurity configuration, varying in a range of few meV per primitive cell. In what follows, we will therefore focus to just one particular configuration of impurities at any given Ru-concentration. 

\begin{table}[]
\caption{In-plane $a$ and out-plane $c$ lattice vector tetragonal cell values used in our calculations, obtained from experiments for BaFe$_{2(1-x)}$Ru$_{2x}$As$_2$ (Ref.~\onlinecite{canfield,ba122struc1,ba122struc2,ba122struc4,ba122kim,baru-acz}), LaFe$_{1-x}$Ru$_x$AsO (Ref.~\onlinecite{la1111struc1,la1111strucF,la1111strucT}) and LaFe$_{1-x}$Ru$_x$AsO$_{0.875}$F$_{0.125}$ (Ref.~\onlinecite{la1111Tc,la1111strucF}).}
\centering
\renewcommand\arraystretch{1.2}
\begin{tabular}{ c c c c c c c } \hline \hline
\multicolumn{2}{c }{} & \multicolumn{5}{c }{Ru-content $x$} \\ 
\multicolumn{2}{c }{} & 0.00 & 0.25 & 0.50 & 0.75 & 1.00 \\ \hline
\multirow{3}{*}{$a$ (\AA)}
& Ba-122 & 3.96 & 4.00 & 4.06 & 4.10 & 4.15 \\
& La-1111 & 4.04 & 4.05 & 4.08 & 4.10 & 4.12 \\
& La-1111F & 4.03 & 4.05 & 4.07 & 4.09 & 4.11 \\
\\
\multirow{3}{*}{$c$ (\AA)}
& Ba-122 & 13.01 & 12.85 & 12.59 & 12.44 & 12.25 \\
& La-1111 & 8.74 & 8.70 & 8.63 & 8.57 & 8.50 \\
& La-1111F & 8.72 & 8.66 & 8.59 & 8.52 & 8.45 \\ \hline \hline
\end{tabular}

\label{params}
\end{table}

In order to avoid spurious effects due to non-perfect correspondence between theoretical predictions and experiments, we use experimental values\cite{canfield,ba122struc1,ba122struc2,ba122struc4,ba122kim,baru-acz,la1111struc1,la1111strucF,la1111strucT,la1111Tc} for the in-plane $a$ and out-plane $c$ lattice parameters to build up the supercells at different Ru-content, as reported in Table~\ref{params} and in Fig.~\ref{fig:structure} (left panels).
On the other hand, in order to save the local effects of the Ru-impurities on the structure, the internal parameters have been obtained performing a full relaxation of the atomic internal positions at any given Ru-concentration.
An energy cutoff of 500~eV is considered to obtain the equilibrium structures, while integration of the irreducible Brillouin zone is performed considering 8$\times$8$\times$3 and 9$\times$9$\times$5 shells for Ba-122 and La-1111(F) respectively, within the Monkhorst and Pack scheme~\cite{Mon76}, until the minimum for the total energy is reached within an uncertainty range of $10^{-3}$~eV.
Once the equilibrium structure is determined, the electronic states are self-consistently converged up to $10^{-4}$~eV total energy difference, using the tetrahedron method, with shells including up to 203 irreducible $k$-points.

At low temperatures both compounds examined are known to be in an orthorhombic antiferromagnetic stripe (AFM2) phase\cite{ba122mossb,afm2,la1111Tc2}, which is kept also at rather large Ru-contents.
We checked that the orthorhombic distortion did not change significantly the results obtained for the tetragonal phase, as far as the energetics of the system and the electronic properties at the Fermi level are concerned\cite{akturk}.
Hence, we focus here on the results obtained for the tetragonal phase only.

On the other hand, a major role in determining the structure of Ba-122 and La-1111(F) compounds is played by the magnetic order of the iron atoms:
the existence of a magnetic order, persisting on a short range scale well below the AFM-paramagnetic phase, has been in fact shown by several experiments\cite{Vilmercati,Mshort}.
This is known to sensibly affect the structural properties of these compounds and in particular the As position with respect to the Fe-planes; as a result, the As-Fe bond results to be strongly dependent on the magnetization of the Fe-atoms.
In order to fully consider these effects, we perform our calculations including a stripe magnetic order (see discussion below) for both compounds and also investigate the magnetic properties as a function of Ru-content.
We perform calculations for non-magnetically ordered cells as well, in order to highlight the differences due to the As position with respect to the stripe phase.

In order to recover the 2-Fe primitive cell band structure and compare the effects introduced by the substitutions directly with ARPES experiments, we use the unfolding procedure, as proposed by Popescu and Zunger\cite{popescuB, popescuL}.
This method has been implemented in the VASP code\cite{VASPunfold,VASPufM} and here briefly presented.
Calculations based on supercell approach, with a unit cell $N$~times larger than the primitive cell, lead to a folded reciprocal space that in most cases makes hard a direct interpretation of the resulting band structure\cite{BerlijnUnf}.
The relation
\begin{equation}
\mathbf{k}+\mathbf{g}=\mathbf{K}+\mathbf{G}
 \label{eq_fold}
\end{equation}
describes the folding of the reciprocal space, mapping a wave vector $\mathbf{K}$ of the reciprocal space of the supercell into $N$ wave vectors $\mathbf{k}$ of the Brillouin zone of the primitive cell (pbz), by means of the reciprocal lattice vectors $\mathbf{g}$ and $\mathbf{G}$ of the primitive and supercell, respectively.
The unfolding method is able to describe the folded energy bands $E(\mathbf{K})$ obtained in the $\mathbf{K}$-vector reciprocal space of the supercell in terms of the $\mathbf{k}$-vector reciprocal space of the primitive cell.
The resulting energy values $E(\mathbf{k})$ can thus be used to visualize the so called effective band structure (EBS), useful for a more clear investigation of the electronic properties of the system.
The EBS is obtained thanks to the projection $P_{\mathbf{K}m}(\mathbf{k})$ of the folded eigenstate $\left | \Psi_{\mathbf{K}m} \right \rangle$ of the supercell into states $\left | \psi_{\mathbf{k}n} \right \rangle$ of the primitive cell (where $m$ and $n$ are band indices).
For eigenstates described by a plane waves basis set, that is our case, the projection $P_{\mathbf{K}m}(\mathbf{k})$, usually referred as Bloch character, can be written in terms of the plane waves coefficients $C_{\mathbf{g}+\mathbf{k},m}$ of the supercell states only, calculated on points connected by reciprocal vectors $\mathbf{g}$ of the primitive cell:
\begin{equation}
P_{\mathbf{K}m}(\mathbf{k}) 
= \sum_n | \left \langle \Psi_{\mathbf{K}m} | \psi_{\mathbf{k} n} \right \rangle | ^2 
= \sum_{\{\mathbf{g}\}} | C_{\mathbf{g}+\mathbf{k},m} | ^2~.
\label{eq:blochch}
\end{equation}
The spectral function $A(\mathbf{k},E)$, which is a useful quantity for a direct comparison with experiments, can be calculated at each $\mathbf{k}$-point as a function of the energy $E$ starting from the $P_{\mathbf{K}m}(\mathbf{k})$ coefficients, as shown by the following equation:
\begin{equation}
A(\mathbf{k},E) = \sum_m P_{\mathbf{K}m}(\mathbf{k})~\delta (E_m - E )~.
\label{eq:specfun}
\end{equation}
The connection between the primitive and the supercell spaces is contained in the transformation matrix~\cite{VASPufM} $\underline{M}$ (invertible and with integer elements) defined by the following equations for the direct ($\underline{\mathbf{A}}$ and $\underline{\mathbf{a}}$) and the reciprocal ($\underline{\mathbf{B}}$ and $\underline{\mathbf{b}}$) lattice vector matrices of the supercell and primitive cell, respectively:
\begin{equation}
\underline{\mathbf{A}} = \underline{M}~\underline{\mathbf{a}}~,
\quad
\underline{\mathbf{B}} = \left ( \underline{M}^{-1} \right )^{T}\underline{\mathbf{b}}~.
\label{eq:AMa}
\end{equation}
Consistently with Eq.~\ref{Avsa}, the transformation matrices used for our calculations are defined as
\begin{equation*}
\underline{M}^{(\textrm{Ba-122})} = 
\left ( \begin{matrix}
  0 & 2 & 2 \\
  2 & 0 & 2 \\
  1 & 1 & 0
 \end{matrix} \right )
~\text{and}~
\underline{M}^{(\textrm{La-1111(F)})} = 
\left ( \begin{matrix}
  2 & 2 & 0 \\
 -2 & 2 & 0 \\
  0 & 0 & 1
 \end{matrix} \right )
\end{equation*}
for Ba-122 and La-1111(F), respectively.
The inclusion into the VASP code\cite{VASPufM} of an automatic tool to map the $\mathbf{K}$-$\mathbf{k}$ reciprocal spaces by means of the $\underline{M}$ matrix accordingly to Eq.~\ref{eq:AMa}, has facilitated the constructions of the Brillouin zones and, consequently, the EBS determination.
In order to be consistent with the size change of the supercells due to different Ru-concentrations, a different primitive cell is defined for each supercell used, keeping constant the given integer elements of the matrix $\underline{M}$.
This provides a set of $\{\mathbf{g}\}$ reciprocal vectors for the calculation of the Bloch character accordingly to Eq.~\ref{eq:blochch}, finally leading to the construction of the EBS in the pbz.

\section{Structural properties}
\label{structure}

\begin{figure}
\begin{center}
\includegraphics[width=0.48\textwidth,angle=0]{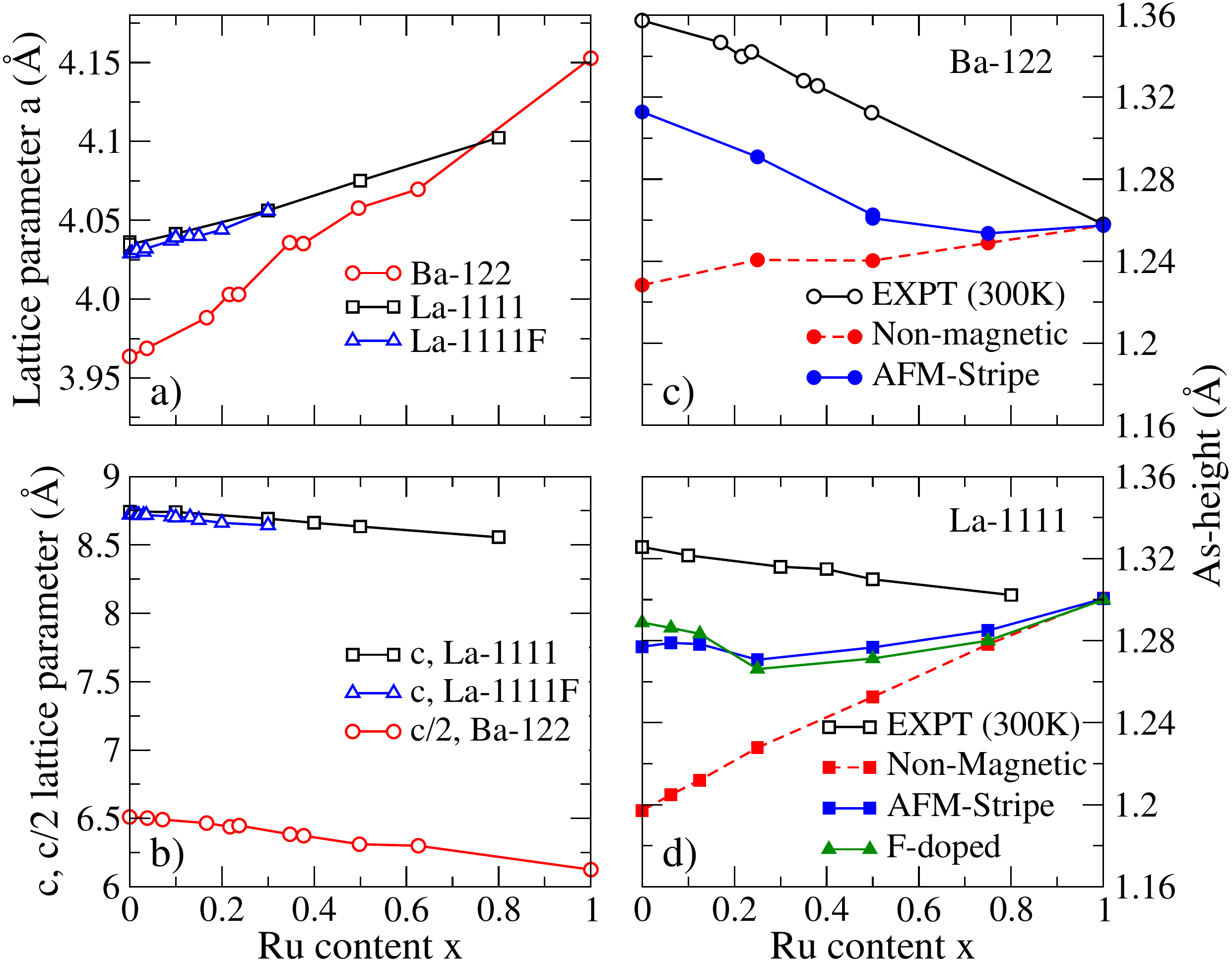}
\caption{(Color online) Left panels: Experimental values (Ref.~\onlinecite{canfield,ba122struc1,ba122struc2,ba122struc4,ba122kim,baru-acz,la1111struc1,la1111strucF,la1111Tc}) of the in-plane $a$ (panel (a)) and out-plane  $c/2$ and $c$ (panel (b)) vectors for Ba-122 (circles), La-1111 (squares) and La1111(F) (triangles) compounds as a function of Ru concentration. Right panels: Experimental values (empty symbols) of the As-height (Ref.~\onlinecite{ba122struc1,baru-acz,la1111struc1,la1111strucF}), with respect to the metallic plane, as a function of Ru concentration for Ba-122 (panel (c)) and La-1111 (panel (d)), compared with {\it ab-initio} non-magnetic (dashed lines) and AFM2-magnetic (continuous lines) calculations. In the case of La1111 results for the stripe-AFM2 F-doped compound are also shown (triangles).}
\label{fig:structure}
\end{center}
\end{figure}

The structural properties will be shown to play a very important role in the electronic states at the Fermi energy level ($E_F$): we therefore discuss in the following how the structure is modified by Ru substitution. Fig.~\ref{fig:structure}  compares the behavior of the lattice constants of Ba-122 with those of La-1111 as a function of Ru-concentration. Due to the larger atomic size of Ru with respect to Fe atoms, the in-plane lattice parameter increases with Ru-content and shows quite sensible variations. However, while the change is lower than 2\% in the La-compound, it is more than twice as big (5\%) in Ba-122. Variations of the $c$ parameter are more similar in the two compounds: 3.8\% and 2.8\% for Ba- and La-compounds, respectively. As a result, in La-1111 the variation is larger for the out-of-plane than for the in-plane parameter, showing that in this case the local internal compression brought by Ru-substitution is more easily adjusted along the out-of-plane direction, since the in-plane size is kept fixed by the more resilient La-O tetrahedral bonds.
In this respect, the Ba-structure appears to be much softer compared to La-1111.

\begin{figure}
\begin{center}
\includegraphics[height=0.3\textwidth,angle=0]{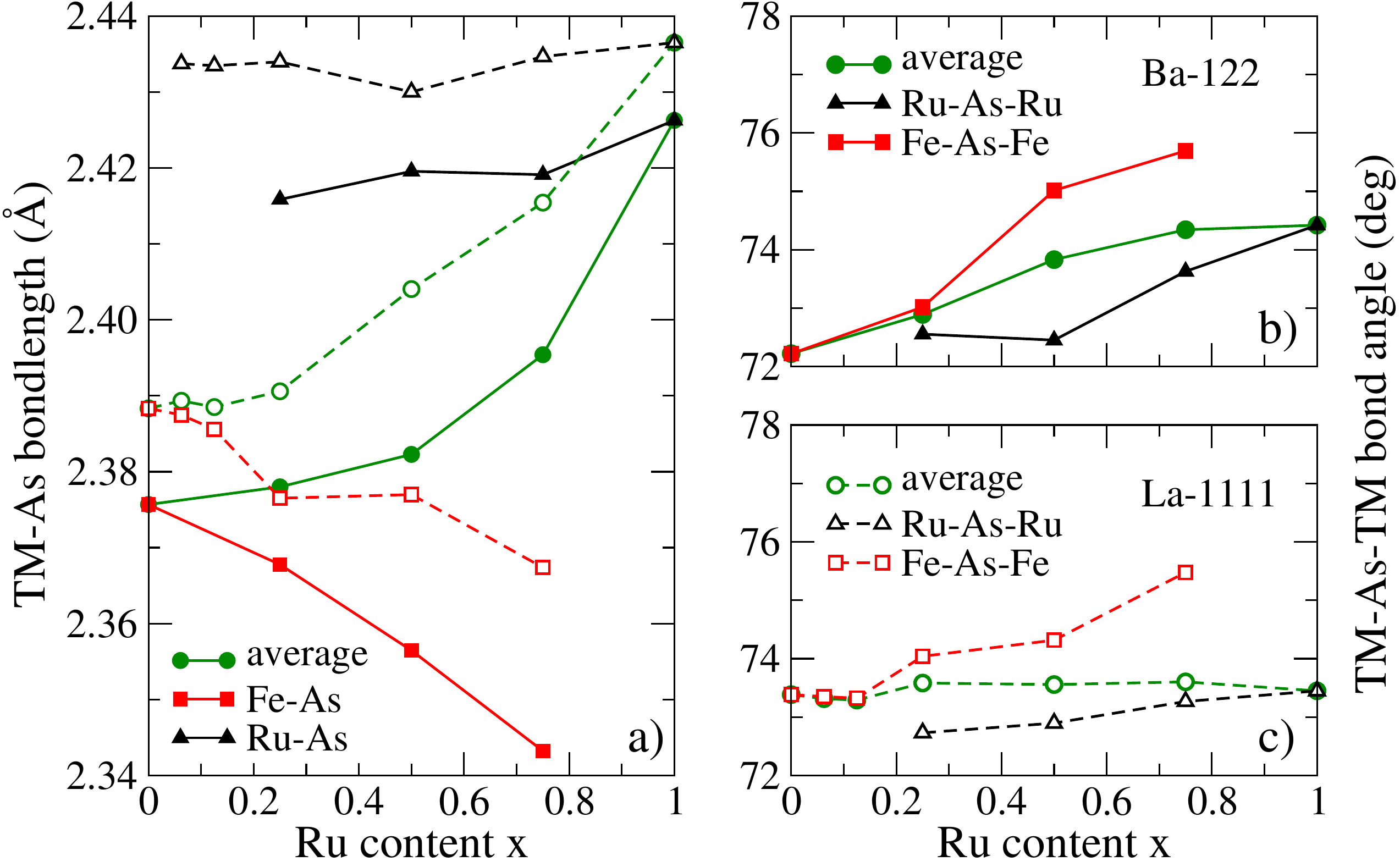}
\caption{(Color online) Panel (a): Fe-As (squares), Ru-As (triangles) and TM-As (circles) bond lengths for Ba-122 (solid lines, filled symbols) and La-1111 (dashed lines, empty symbols) as a function of Ru concentration. Right panels: Fe-As-Fe (squares), Ru-As-Ru (triangles) and TM-As-TM (circles) bond angle for Ba-122 (panel (b)) and La-1111 (panel (c)) as a function of Ru concentration.}
\label{fig:astm}
\end{center}
\end{figure}

Fig.~\ref{fig:structure} (right panels) shows {\it ab initio} results for the As-height $z_{\rm As}$ from the Fe-plane fully relaxed in the non-magnetic and in the stripe-AFM2 magnetic phase, compared with experimental values at room temperature (RT)\cite{la1111struc1,ba122struc1,baru-acz,la1111strucF}. The non-magnetic calculations show a behavior of the $z_{\rm As}$ internal parameter values completely different with respect to the experiment. As systematically studied in the literature\cite{DFTvsEXP}, we find that the magnetic ordered phase better reproduce the experimental behavior, even for the RT-structures, where long range AFM2 order is known not to survive: this agrees with experiments which find  non-vanishing local fluctuating magnetic moments, giving raise to a long-range paramagnetic structure\cite{Vilmercati}.
In the case of Ba-122, the As-height  shows a linear reduction of 0.05~\AA \ going from $x=0$ to $x=0.5$, nicely  following the same slope as in experiment.
In La-1111, both pure and F-doped, this same parameter shows a much slower variation as a function of dopant concentration, indicating a larger stiffness of the La-1111 structure to adjust internal strain effects, when compared to Ba-122.
In addition, at large Ru contents ($x \geq 0.8$), the experimental data remarkably tend to the DFT values.
We recall here that the As distance from the Fe-plane is a structural parameter that has been often related to the superconducting properties~\cite{zAs-Tc}, as it is strongly linked to the electronic correlations characterizing the d-orbitals involved in the As-TM bond and to the occurrence of magnetic fluctuations.
For these same reasons, the values obtained within DFT-based approaches always underestimate the experimental values. The closer agreement between theory and experiments found here at large Ru-content, when the magnetic order disappears (see discussion below), would certainly indicate a reduced relevance of correlation effects in the Ru-rich structures.

 Fig.~\ref{fig:astm} (left panel) shows the calculated values of the relevant bond lengths involving the TM and As atoms as a function of Ru-concentration.
We find that in both compounds the Ru-As bond is kept quite constant even at rather small Ru-content; on the other hand, the Fe-As bond is affected by larger changes which are definitely more evident in the Ba-based compound.
The average of the bond-lengths is also shown, providing a good agreement with experiments\cite{ba122struc1}. The same trend is found for the bond-angle (see right panels of Fig.~\ref{fig:astm}): the La-1111 compound shows a remarkable robustness of the Ru-As-Ru bond angle, related to an essentially constant Ru-As bond-length. 
Thus, local distortions are definitely more relevant in Ba-122 where the overall structure as well as the internal parameters show sensible deviations from the pure compound.
Since the Fe-As local bond is quite deeply affected by Ru-substitution, it is reasonable to  expect that Ru would induce quite large disorder effects that are going to be more relevant in Ba-122 than in La-1111 compounds. 

\begin{figure}
\begin{center}
\begin{tabular}{c}
\includegraphics[width=0.39\textwidth,angle=0]{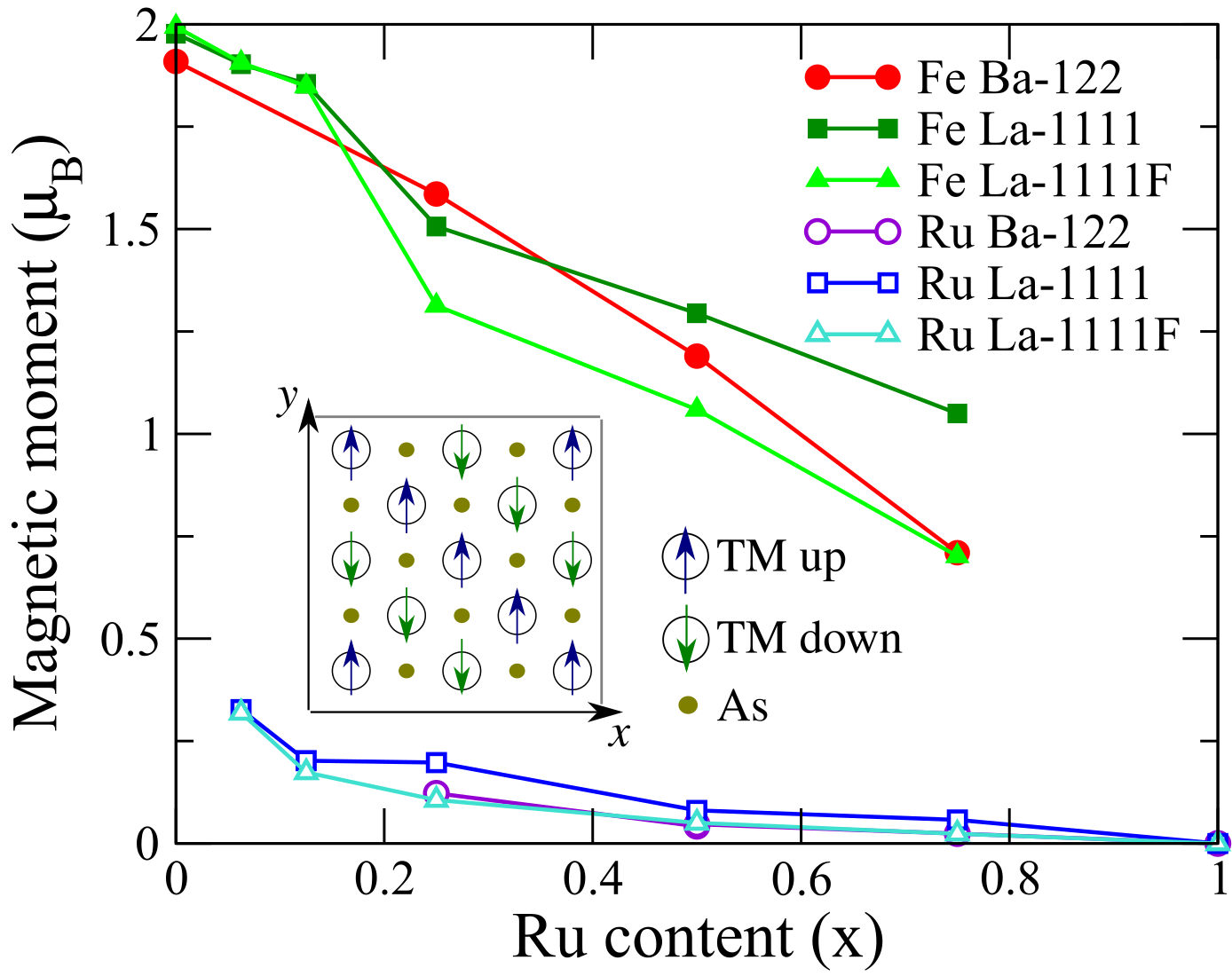}  \\
\end{tabular}
\caption{(Color online) Calculated local magnetic moment of Fe (filled symbols) and Ru (empty symbols) in Ba-122 (circles), La-1111 (squares) and La-1111F (triangles) as function of Ru concentration. The inset sketches the ordering of the magnetic moments on the TM sites, in the stripe (AFM2) phase and the orientation of the metallic plane as considered in our calculations.}
\label{fig:magn}
\end{center}
\end{figure}

To complete this section, we briefly address the magnetic properties of the two compounds as a function of Ru-content. Fig.~\ref{fig:magn} shows the calculated magnetic moment on the Fe and Ru sites obtained for the compounds studied: Ba-122, La-1111 and La-1111F.  We see  that the magnitude of the magnetic moment on the Fe and Ru sites is not strongly dependent on the compound considered and decreases as the Ru-content increases, in agreement with experiments (despite discrepancies in the absolute values, well known in literature\cite{toschi,toschiI,ba122m,la1111m,la1111m2}).
In addition, the decrease of the local magnetic moment on the Fe sites mimics very well the behavior of the As-height shown in Fig.~\ref{fig:astm} (left panel) for Ba-122 and La-1111. This is once again a confirmation of the close relationship between the two quantities. The magnetic moment on the Ru-site is seen to be much smaller than the one on the Fe-sites and fades quickly away as the Ru content increases. This is a clear indication of the more delocalized Ru bonds (see discussion on the density of states in Sect.~\ref{DOS}).

\section{Electronic properties}
\label{electronic}
\subsection{Band structure at the Fermi level: Ba-122}
\label{electBa}

\begin{figure*}
\begin{center}
\includegraphics[width=1.0\textwidth,angle=0]{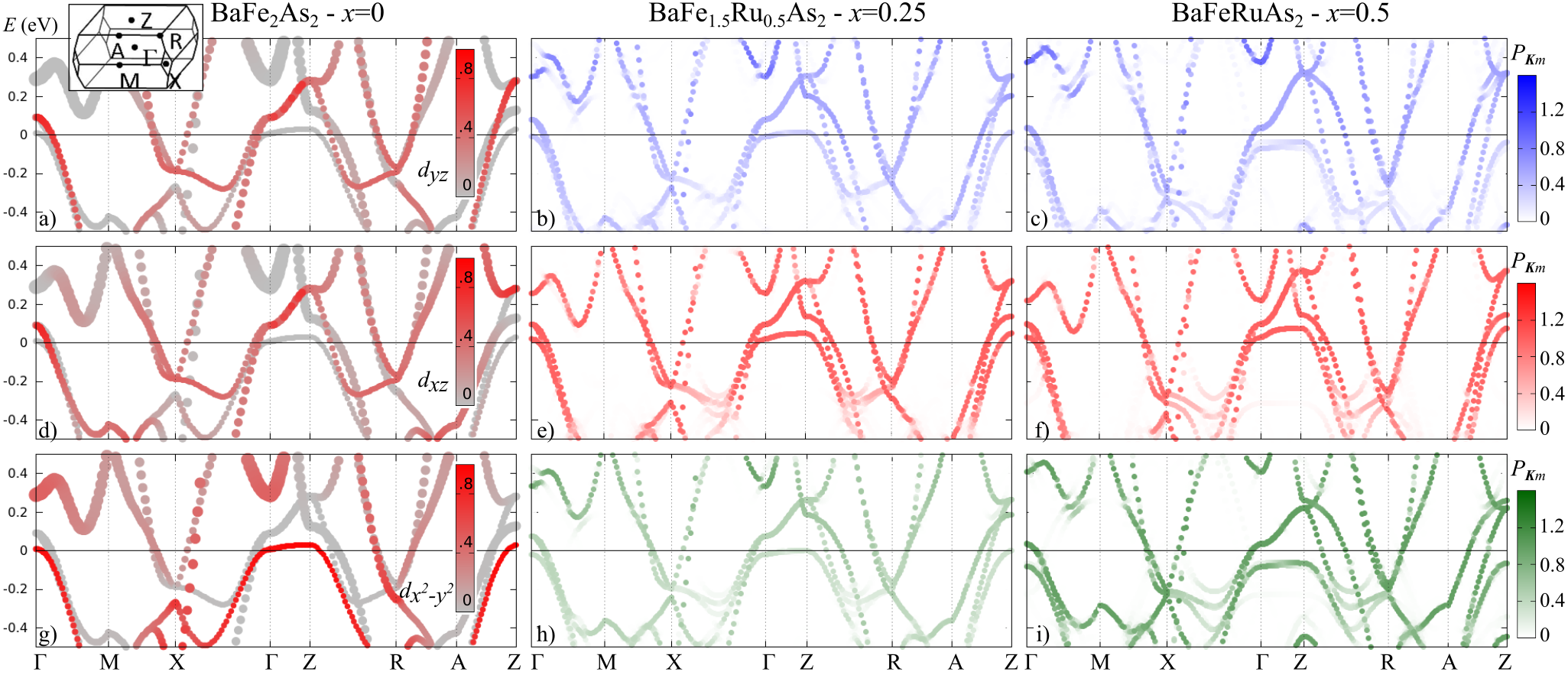} 
\caption{(Color online) Effective band structure (EBS) for Ba-122 as function of Ru concentration ($x= 0.0, 0.25, 0.5$), unfolded on the 2 Fe-bct primitive cell. Panels (a), (d) and (g) show the band structure of  pure Ba-122  with  highlighted the projected orbital character  ($d_{yz}$, $d_{xz}$, $d_{x^2-y^2}$, respectively); here, the size of the data points is proportional to the amount of Bloch character. Panels (b) and (c) show the band structure for the fully relaxed structure (standard calculation) at $x=0.25$ and $x=0.5$ Ru-content, respectively. Panels (e) and (f) report the band structure of the Ru-doped compound (at $x=0.25$ and $x=0.5$ Ru-content, respectively) in the structure of pure Ba-122. Finally, panels (h) and (j) show the band structure obtained for pure Ba-122 in the fully relaxed doped structures (at $x=0.25$ and $x=0.5$ Ru-content, respectively). The inset sketches the high symmetry points in the pbz.}
\label{fig:ba122bands}
\end{center}
\end{figure*}

\begin{figure}
\begin{center}
\includegraphics[width=0.45\textwidth,angle=0]{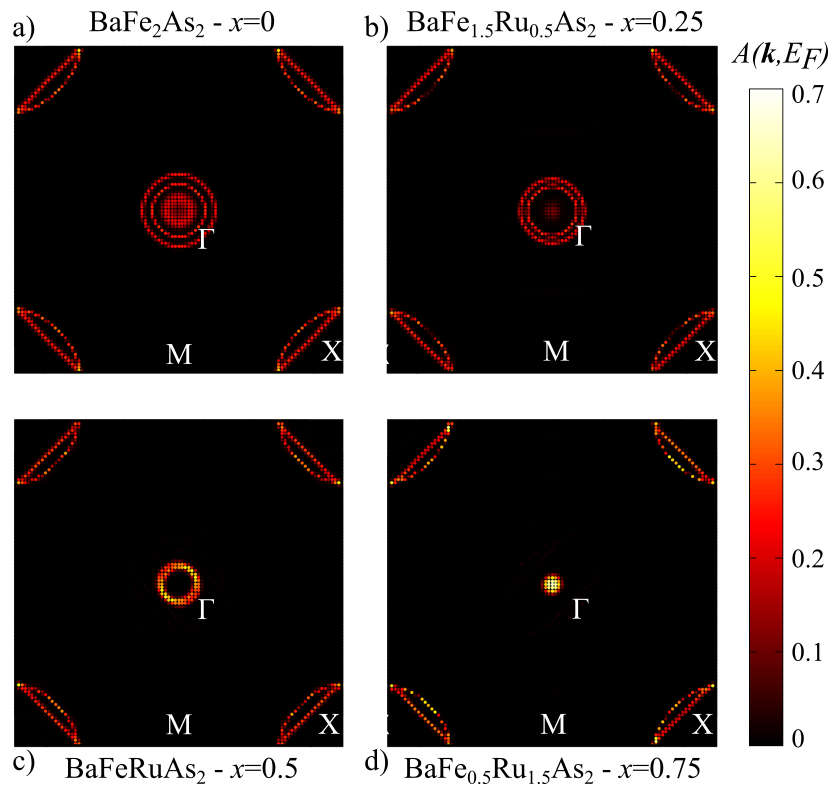}
\caption{(Color online) Fermi surface for Ba-122 as function of Ru concentration, obtained by the standard calculation at the given Ru-content: $x = 0.0, 0.25, 0.5, 0.75$ (panel (a),(b), (c) and (d), respectively). The color gradient represents the value of the spectral functional $A(\mathbf{k},E_F)$ calculated accordingly to Eq.~\ref{eq:specfun}.}
\label{fig:ba122FS}
\end{center}
\end{figure}

\begin{figure*}
\begin{center}
\includegraphics[width=1.0\textwidth,angle=0]{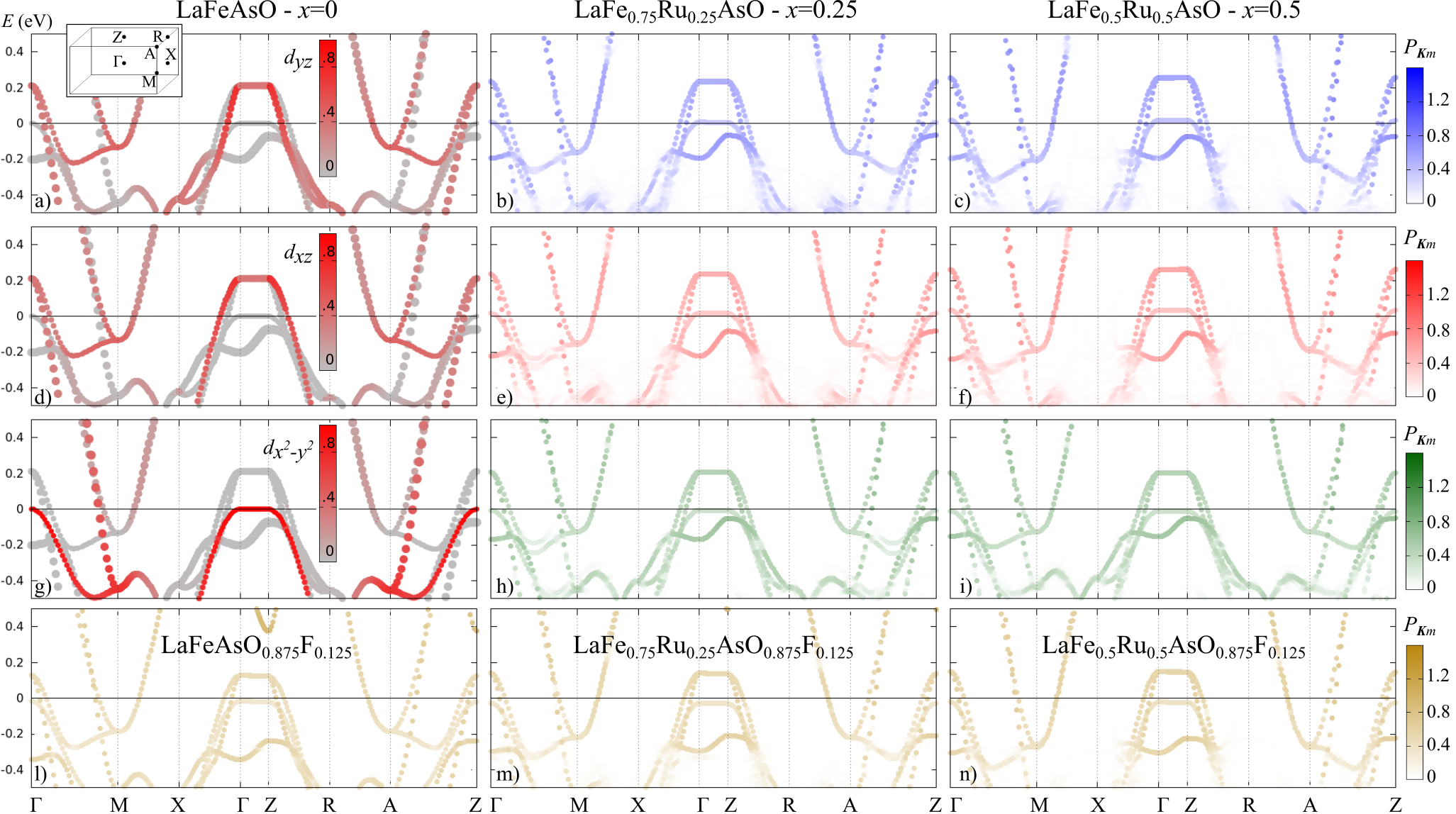}
\caption{(Color online) Effective band structure (EBS) for La-1111 and La-1111F as function of Ru concentration  ($x= 0.0, 0.25, 0.5$), unfolded on the 2 Fe primitive cell. Panels (a), (d) and (g) show the  band structure of the pure La-1111 compound decorated with the orbital character ($d_{yz}$, $d_{xz}$, $d_{x^2-y^2}$, respectively); here the  size of the data points is proportional to the amount of Bloch character. Panels (b) and (c) show the band structure for the fully relaxed structure (standard calculation) of La-1111 at $x=0.25$ and $x=0.5$ Ru-content, respectively. Panels (e) and (f) report the band structure of  Ru-doped La-1111  (at $x=0.25$ and $x=0.5$ Ru-content, respectively) in the structure of pure La-1111. Panels (h) and (i) show the band structure obtained for pure La-11111 in the fully relaxed doped structures (at $x=0.25$ and $x=0.5$ Ru-content, respectively). Finally, panels (l), (m) and (n) show the band structure for the fully relaxed structure (standard calculation) of F-doped La-1111 compound at $x=0.0$, $x=0.25$ and $x=0.5$ Ru-content, respectively. The inset sketches the high symmetry points in the pbz.}
\label{fig:la1111bands}
\end{center}
\end{figure*}

We start the investigation of the electronic states at the Fermi level, discussing the electronic band structure of pure Ba-122  in its equilibrium structure, as shown in Fig.~\ref{fig:ba122bands}. Here, the panels on the left column highlight the $d_{yz}$ (panel (a)), $d_{xz}$ (panel(d)) and $d_{x^2-y^2}$ (panel(g)) orbital characters. Let us recall that, in the geometry considered (see inset in Fig.~\ref{fig:magn}), the first two states ($d_{yz}$ and $d_{xz}$) lay along the Fe next-nearest-neighbors and involve the Fe-As bonds, while the third one ($d_{x^2-y^2}$), which is also the lowest one in energy, points towards the Fe-Fe next-nearest-neighbors direction as well but lays in the TM-plane. As expected, the $d_{xy}$ orbital (not shown), involved in the Fe-Fe nearest-neighbor bonds, appears at very deep energy. In agreement with previous calculations\cite{ba122unfold}, these three $d_{yz}$, $d_{xz}$ and $d_{x^2-y^2}$ bands cross the Fermi level, forming hole pockets at the zone centers $\Gamma$, $Z$ and  electron pockets at the zone corners $X$, $R$. The lower band, being essentially localized in the Fe-planes, has a very small dispersion along the $k_z$-direction and gives rise to an essentially cylindrical sheet along the $\Gamma-Z$ direction, as widely discussed in the literature\cite{ba122expFS}.

We now consider how the states at the Fermi level change upon Ru substitution.
Fig.~\ref{fig:ba122bands} shows the band structure, unfolded in the 2 Fe-bct cell of the pure compound, calculated at different Ru-content ($x= 0.25, 0.5$, panels (b) and (c), respectively), fully relaxing the internal parameters at each Ru-concentration.
In the following, we will refer to these structures as 'standard structures'.
The most evident changes induced by Ru-doping involve the hole pockets at $\Gamma$ (more clearly visible along the $\Gamma-M$ line) and at $Z$: as the Ru concentration increases, the smallest pocket ($d_{x^2-y^2}$ character) progressively closes and becomes fully occupied at concentrations larger than $x=0.25$.
As a consequence, the inner cylindrical hole-sheet along the $\Gamma-Z$ line disappears completely.
The external $d_{xz}$ and $d_{yz}$ bands close around the zone center ($k_z$ = 0), while they widen on the top plane ($k_z$ = $\pi/2c$) enlarging their Fermi vectors, bringing the band-top at higher energy with respect to pure Ba-122.
Other minor changes affect the electron pockets at $X$ and $R$, with a band-bottom becoming deeper in energy as the Ru-content is increased, while the Fermi vectors are kept constant.
The changes at Fermi surface (FS) induced by Ru-substitution are shown in Fig.~\ref{fig:ba122FS} where the FS evolution as a function of Ru-content is projected on the basal plane ($k_z=0$).
Panels (a) to (d) of Fig.~\ref{fig:ba122FS} show the progressive vanishing of the hole pockets at the zone center and the essential invariance of the electron-like sheets at the $X$-point upon Ru doping.

Neglecting the well known problems related to renormalization of the theoretical band-width due to electronic correlation effects, the comparison with ARPES measurements\cite{ba122expFS,brouet2010} gives a qualitative agreement for the $d_{xz}$ and $d_{yz}$ bands, while the behavior of the $d_{x^2-y^2}$ is completely different since this band is experimentally found to be only slightly affected by Ru substitution, and to cross the Fermi level even at large $x$ concentration.
This discrepancy, common to all DFT calculations\cite{ba122unfold,ba122DOS}, can find an explanation by performing a 'computational experiment': we  take the  Ru-contents just discussed ($x = 0.25, 0.5$) and change the structure, forcing the doped compound into the pure Ba-122 structure.
The resulting band structure is shown in the central row of Fig.~\ref{fig:ba122bands} ($x=0.25, 0.5$, panels (e) and (f), respectively).
Several noticeable differences stem from the comparison of the central (e) and right (f) panels of the second row with the corresponding (b) and (c) panels reported in the upper row of Fig.~\ref{fig:ba122bands}: the most affected state is the $d_{x^2-y^2}$ hole pocket at the zone center which now grows in size as the Ru content increases, slightly enlarging the Fermi vector.
The cylindrical inner hole-sheet along the $\Gamma-Z$ becomes larger and acquires a 3-dimensional dispersion; at the same time, the Fermi vectors of the electron pockets at $X$ and $R$ are kept almost constant, while the bottom of these pockets move to deeper energies.
Thus, Ru-potential itself sensibly changes the Fermi surface with respect to the standard case, while the state energy distribution is overall kept, as further discussed in Sect.~\ref{DOS}.

As a next step, we proceed further with our 'computational experiment': we fix the fully relaxed structures corresponding at each Ru-concentration  ({\it i.e.} the same structures considered in Fig.~\ref{fig:ba122bands} panels (b) and (c)) and dress them with the pure Ba-122 compound at full Fe content ({\it i.e.} without Ru-doping, $x=0$).
The resulting band structures are shown in the bottom row (panels (h) and (i)).
Comparing the corresponding doped structures (standard structures in panel (b) and (c)), we find that the main features at the Fermi level are very similar: the closure of the $d_{x^2-y^2}$ hole-like pocket is kept and the effects on the Fermi vectors of the different orbitals are very well mimicked, as the virtual Ru-content is progressively increased.
On the contrary, the band-bottom of the electron pockets at $X$ and $R$ remains fixed, instead of moving towards higher binding energies upon Ru substitution as found for the standard structures of panels (b) and (c).

Then, it is possible to conclude that the changes induced by Ru-substitution at the Fermi level are mainly due to a purely structural effect:
the larger ionic size of Ru increases the in-plane lattice constant (see Fig.~\ref{fig:structure}) and the Fermi surface changes its shape shrinking the d$_{x^2-y^2}$ hole-like pocket, regardless the real atomic potential occupying each single Fe-site;
the mechanism is very similar to what observed in more complex $RE_4Fe_2As_2Te_{1-x}O_4$ (42214) compounds upon change of the rare-earth element\cite{bucci}.
In addition, also the behavior of the Fermi $k_F$ vectors corresponding to the $d_{xz}$ and $d_{yz}$ bands (see later discussion Sect.~\ref{subs:pocket}) obtained in the standard structure calculation, is found to be better described by the structures relaxed at the proper Ru concentration but containing only Fe atoms, rather than by the pure structure enriched with the proper amount of Ru impurities.
As a result, we can conclude that the states at the Fermi level are essentially sensitive more to structural  than to chemical effects ({\it i.e.} Ru impurity potential). 
A quantitative characterization of the role played by these two effects is given in Sect.~\ref{subs:pocket}, together with a comparison with the La-1111 compound.

Consistently with this view, the discrepancy between the calculated and the experimental behavior of the $d_{x^2-y^2}$ hole-pocket at $\Gamma$ and $Z$ upon Ru substitution finds a clear explanation.
In fact, as evident from our 'computational experiment' (Fig.~\ref{fig:ba122bands}), the $d_{x^2-y^2}$ band strongly depends on the $z_{\rm As}$ internal parameter.
However, due to well-known shortcoming of DFT functionals in treating correlation effects, the calculated value of this parameter is always lower than experiments (see Fig.~\ref{fig:structure}).
Evidently, the GGA approximation works with different degree of success in determining the appropriated relaxed internal parameters depending on the amount of Ru content that influence the degree of electronic correlation.
Thus, the premature closing of the DFT $d_{x^2-y^2}$ hole-pocket, comes together with the underestimated $z_{\rm As}$ internal parameter.
In our 'computational experiment' (see panels (e) and (f) in Fig.~\ref{fig:ba122bands}), the structure is fixed, and in particular the $z_{\rm As}$ parameter keeps its largest calculated value of 1.31~\AA, roughly corresponding to the experimental value for a concentration of $x=0.5$: in this case the same $d_{x^2-y^2}$ band better agrees with ARPES experiments\cite{ba122expFS} up to large Ru content.
Once again, the accurate determination of the structure seems to be fundamental to achieve a satisfying description of the electronic properties of Ba-122 at the Fermi level.

\subsection{Band structure at the Fermi level: La-1111}
\label{electLa}
We now proceed performing the same kind of experiment on the La-1111(F) compounds at the same Ru concentrations considered for Ba-122.
Fig.~\ref{fig:la1111bands}  (left column: panels (a), (d), (g), (l)), reports the band structure close to the Fermi level  for the La-1111 pure compound  with the Fe-$d_{yz}$  (panel (a)), the Fe-$d_{xz}$  (panel (d)) and Fe-$d_{x^2-y^2}$  (panel (g)) characters highlighted and for the F-doped La-1111 compound (panel (l)).
Comparing the band structure of the superconducting compound (panel (l)) with the parent compound (panel (a)), the effect of Fluorine doping stems out clearly: the added electrons change the size of the topmost hole-pockets at the zone center without changing their 2D character\cite{hosonoRev} (note the absence of any dispersion along the $\Gamma-Z$ line), and at the same time increase the partial filling of the electron pockets along the zone corner ($M$ and $A$ points), leaving unchanged the lower $d_{x^2-y^2}$ states. The electrons coming from the F-atoms substituting on the LaO-planes go into the tetrahedral Fe-As bonds, shifting almost rigidly the Fermi level towards higher energies.

We now move to Ru-doping: at $x=0.25$ very little changes are hardly visible in the fully relaxed structures and, as we further increase the Ru-content, a small shift towards lower energies of the electron pockets at $M$ and $A$ points can be detected. All the other states are not affected in both cases, the pure and F-doped La-1111 compounds (Fig.~\ref{fig:la1111bands} panels (b),(c) and (m),(n), respectively).

We then proceed with our 'computational experiment', similarly to the Ba-122 case: Fig.~\ref{fig:la1111bands} shows the band structure of the La-1111 compound at the same Ru-content as before ($x=0.25$, and $x=0.5$, panels (e) and (f), respectively) constrained in the pure La-1111 structure. We can easily see that the band structure is not changed with respect to the La-1111 'standard structures' (panels (b) and (c)); only a very small variation can be appreciated at quite large Ru-content ($x=0.5$) in the  $d_{x^2-y^2}$ state at the Fermi level: comparing with panel (c) (corresponding to the same concentration but in the fully relaxed structure) it is possible to appreciate a small shift upwards of the 2D-inner hole sheet which now forms a tiny pocket (more evident along the $\Gamma$-$Z$ line).

We finally move to the third step of the experiment and consider the fully relaxed Ru-doped La-1111 structure and dress it with the pure compound ({\it i.e.}  full Fe-content in the $x= 0.25, 0.5$ structures),  Fig.~\ref{fig:la1111bands} (panels (h), (i)). Also in this latter case we find that the states at the Fermi level are very similar to those of the doped compound in 'standard structures': in particular, the size of the hole-pockets at the zone center appears to be unchanged, while a small shift towards higher energies is found for the electron-pockets at the corner points $M$ and $A$. 
 
Thus, the pure structural effect as well as the chemical effect are seen to play a very minor role in this case concerning the topology of the Fermi surface. The fact that the impurity potential (chemical effect) is not affecting the states at the Fermi level agrees with what found in the Ba-122 case; at the same time the structure, that was shown to sensibly affect the Fermi topology and the states occupation in the Ba-case, is now not playing any relevant role since the La-1111 structure is not dramatically changed by Ru-substitution (as discussed in Sect.~\ref{structure}).

\begin{figure}
\begin{center}
\includegraphics[width=0.47\textwidth,angle=0]{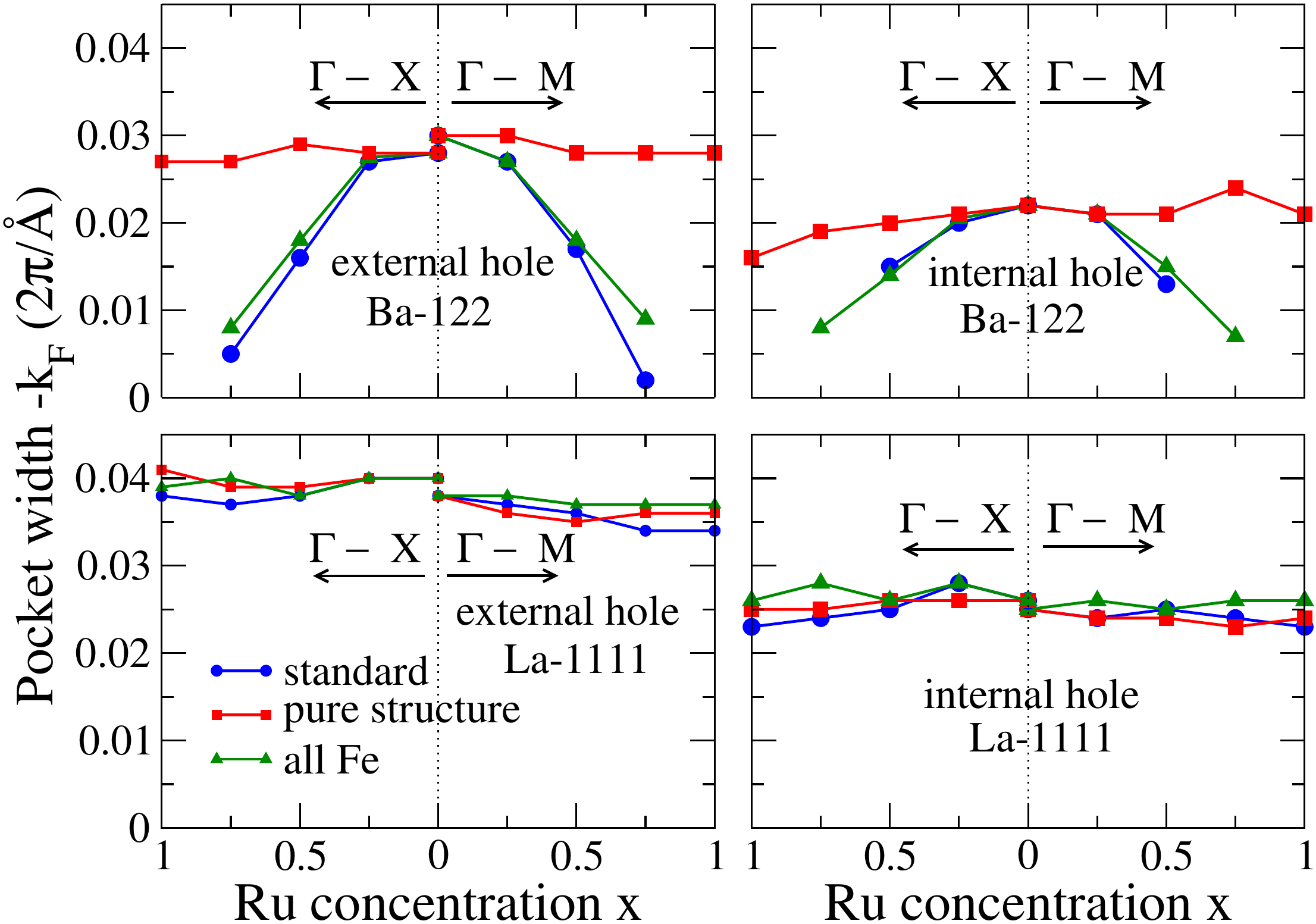} \\
\caption{(Color online) Size of the internal and external $d_{xz}$ and $d_{yz}$ hole pockets ({\it i.e.} the Fermi vectors $k_F$) for Ba-122 and La-1111 as function of Ru concentration in the $\Gamma-X$ and $\Gamma-M$ directions. Circles represent the results obtained  using structures with fully relaxed internal parameters (standard structures); the lines with squares have been obtained using the structures of the pure compounds dressed with the proper amount of Ru atoms; the lines with triangles represent the result obtained with relaxed Ru-doped structures dressed entirely with Fe atoms at each  TM site.}
\label{fig:pockets}
\end{center}
\end{figure}

\subsection{Size evolution of hole and electron pockets}
\label{subs:pocket}
A more quantitative proof of the results just discussed in Sect.~\ref{electBa} and~\ref{electLa}, in addition to Fig.~\ref{fig:ba122bands}, \ref{fig:ba122FS} and \ref{fig:la1111bands}, can be achieved by looking at the evolution of the size of the hole pockets at the zone center in both Ba-122 and La-1111 compounds, as a function of Ru-content.
Fig.~\ref {fig:pockets} shows the Fermi vector $k_F$ ({\it i.e.} the distance from the $\Gamma$ point to the k-point where each $d_{xz}$ and $d_{yz}$ hole-pocket sheet crosses the Fermi level, along the $\Gamma$-$X$ and $\Gamma$-$M$ directions) as a function of Ru-content and for all the different cases considered, namely: fully relaxed structure (standard calculation), structure of the pure compound (pure structure with proper amount of Ru-doping) and full Fe-content ($x=0$ in the fully relaxed structure), each at the corresponding Ru-content.
First of all, the isovalent Ru-substitution in the Ba-122 system fills the hole-pockets at $\Gamma$ (with a compensating mechanism at $Z$ together with a deepening of the binding energies at $X$ and $R$, as visible from the band structure in Fig.~\ref{fig:ba122bands}).
Furthermore, we notice that the hole size is practically unchanged in the La-1111 compound, independently on the structure or impurity potential considered.
Finally, we find that the calculations for the fully relaxed structure with 100\% Fe-content, very well reproduce the correct size of both the $d_{xz}$ and $d_{yz}$ inner and external hole pockets in  Ba-122, even at quite large Ru-content (up to 50\%, at least); on the other hand, considering the proper Ru content in the pure Ba-122 structure leads to a completely different description of the topology at Fermi, with practically unchanged hole pockets.

Thus, we can state that, as far as the states at the Fermi level are concerned, a major role in changing the state occupancy at $E_F$ is played by the structural changes brought by the impurity, rather than by the impurity potential itself\cite{Merz}.

\begin{figure}
\begin{center}
\begin{tabular}{cc}
\includegraphics[width=0.24\textwidth,angle=0]{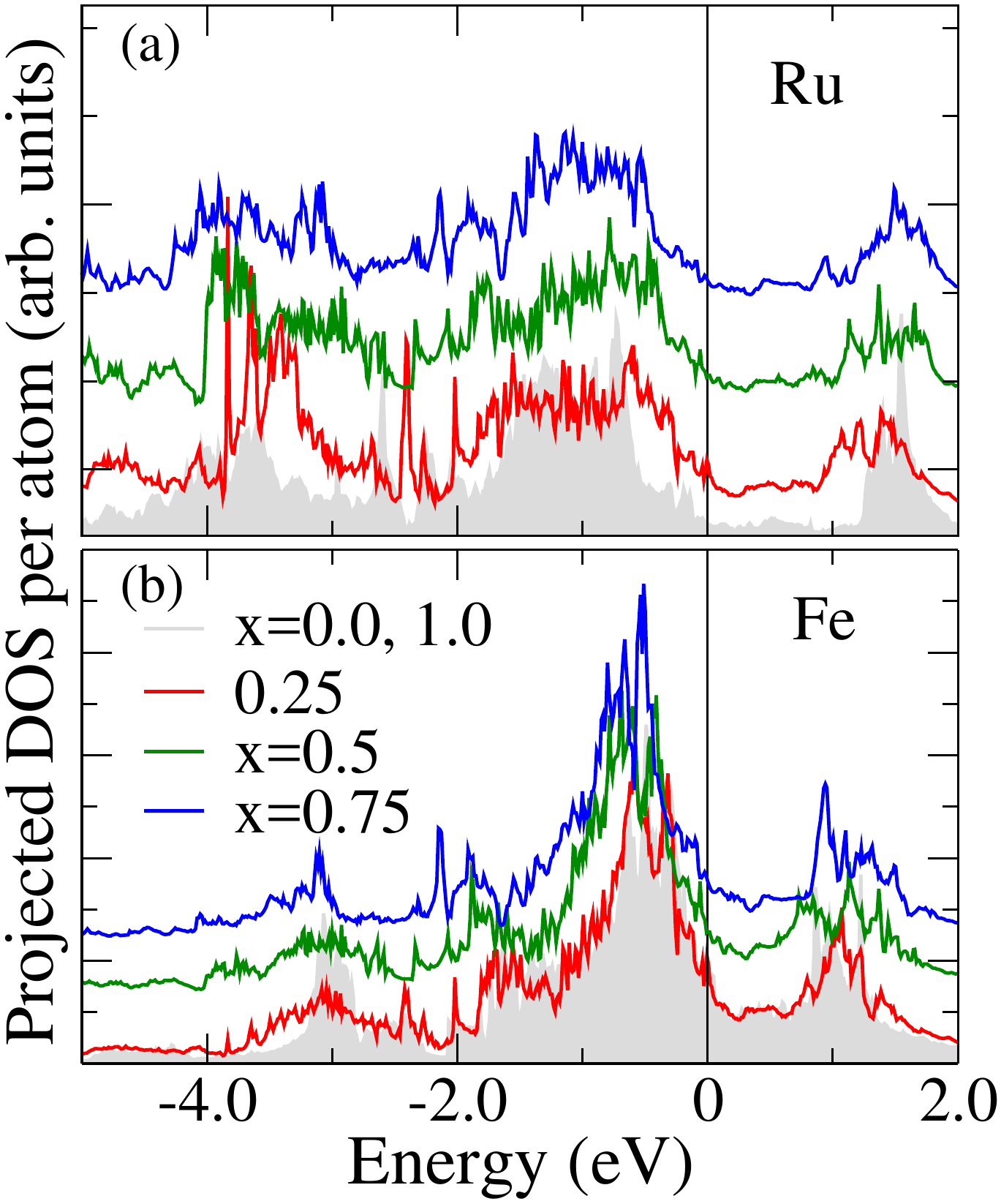} & 
\includegraphics[width=0.24\textwidth,angle=0]{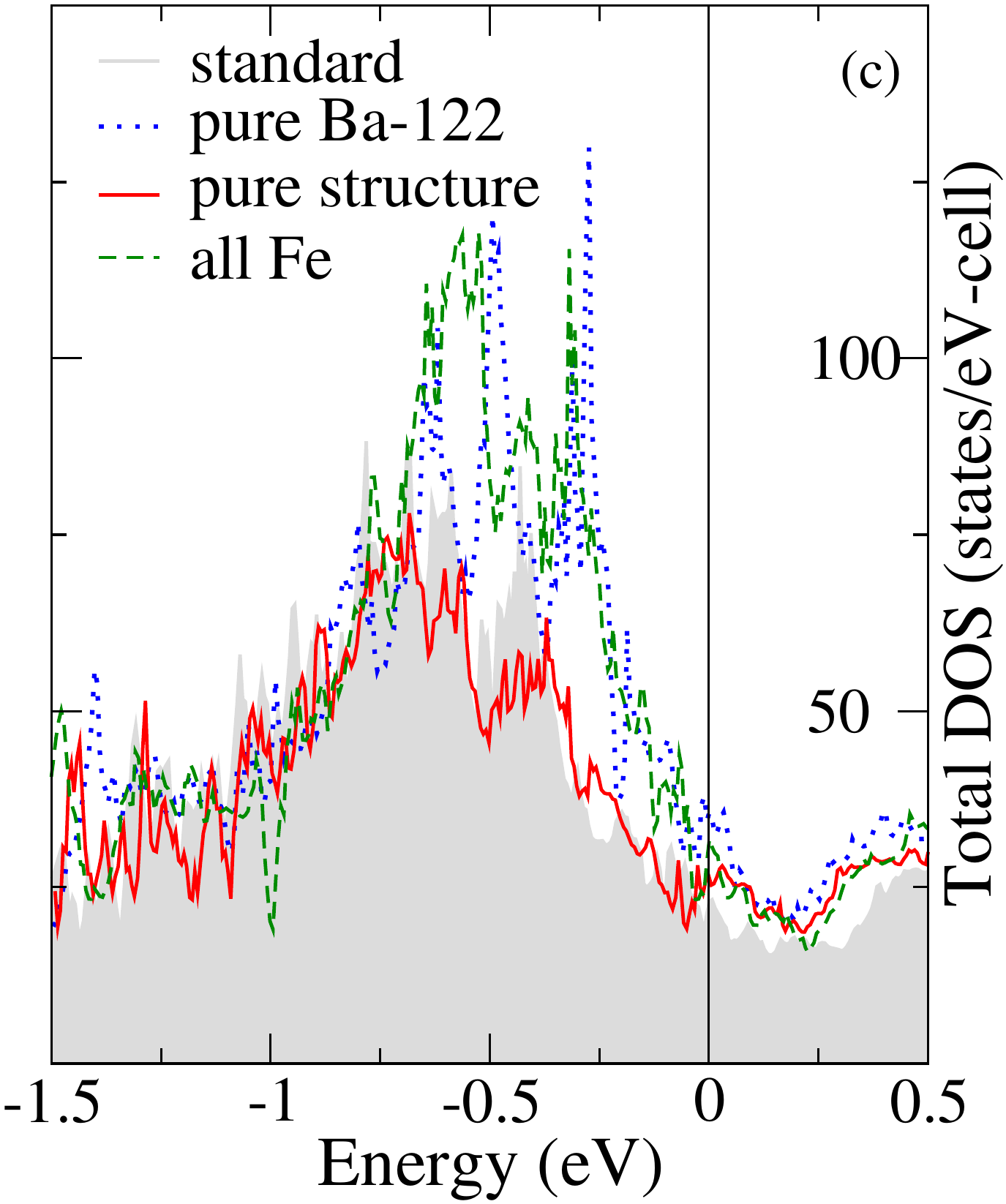} \\
\end{tabular}
\caption{(Color online) Atom projected DOS for Ba-122 on the Ru (panel (a)) and Fe (panel (b)) sites at different Ru-content ($x= 0, 0.25, 0.5, 0.75, 1$), in arbitrary units. The partial density of states corresponding to different concentrations is arbitrarily shifted on the $y$-axes to allow comparison. Panel (c): Total DOS for Ba-122 at $x=0.5$ calculated for the fully optimized (gray filled curve), all Fe (relaxed structure dressed up with Fe atoms only, dashed line) and pure (pure structure dressed up with 50\% Ru doping, continuous line) structures, and compared with the fully optimized pure $x=0$ Ba-122 case (dotted line).}
\label{fig:ba122DOS}
\end{center}
\end{figure}

\section{Impurity potential, binding properties and density of states}
\label{DOS}

As previously discussed, we found the electronic band structure at the Fermi level to be strongly dependent on the structural parameters.
Consequently, since the structures of the Ba-122 and La-1111(F) compounds are affected by Ru substitution to a different degree, the effects on the band structures are different:
Ru is varying the hole-pocket size at $\Gamma$ and $Z$ in Ba-122 while is not changing to any appreciable extent the Fermi topology in La-1111, as just shown in the previous section. 
Therefore, now the question arises: what is the direct role, if any, of the impurity potential and how does it  affect the overall electronic properties of each compound?

To answer this question we report in Fig.~\ref{fig:ba122DOS} (left panels) the density of states (DOS) projected on the Fe and Ru sites of the Ba-122 compound as obtained from the fully relaxed structures at the various Ru-concentrations considered.
Analogous results are obtained for the La-1111 compound (not shown here).
As expected, being the Ru-4$d$ shell more delocalized than the Fe-3$d$ shell, the energy distribution of the occupied states of Ru is much less pronounced at energies close to the Fermi level with respect to Fe.
The major peak of the Fe-DOS moves towards higher binding energies as the Ru-content increases.
Furthermore, the delocalized character of the Ru-orbitals gives also rise to a wider band (centered around -4~eV) extending at higher binding energies and broadening the hybridization with the As-$p$ states (not shown) at energies lower than -4 eV (as also recently found for Co doping\cite{ba122expCo}).
This also shifts the Fe-peak from -3~eV towards higher binding energies, making the Fe-DOS less and less structured as Ru content increases.

Fig.~\ref{fig:ba122DOS} (right panel) shows the effect of the Ru-substitution on the density of states at energies close to the Fermi level, comparing the DOS calculated for different structures: namely, the standard calculation ({\it i.e.} the fully relaxed Ru-doped structure at $x=0.5$), the un-doped compound at the fully relaxed structure ({\it i.e.} the $x=0$ compound in the $x=0.5$ structure), the doped system in the structure of the pure compound ({\it i.e.} the $x=0.5$ compound constrained in the structure of the pure compound), and, finally, the pure Ba-122 compound in its own fully relaxed structure.
Unrevealing the effect of the impurity potential is not trivial, though an overall behavior appears clearly.
The curve representing the fully relaxed doped compound at $x=0.5$ better compares with the curve representing the same compound but in the pure structure, rather than with the DOS of the other two cases (all Fe in the relaxed $x=0.5$ structure, and pure Ba-122 standard structure, which, by the way are quite similar to each other), giving larger deviations even if the all Fe case is constrained in the same structure.
Thus, as far as the overall energy distribution of the states is considered, the major role seems to be played by the impurity potential rather than by the local structural details ({\it i.e.} the opposite of what we found for the Fermi vectors in Sect.~\ref{electronic}).

Therefore, while structural changes modify the Fermi surface, the impurity potential shifts the TM-As hybridization to lower energies upon Ru substitution.
We stress here that only the combination of both effects lead to a proper description of the band structure, that otherwise is difficult to achieve using other methods (such as virtual crystal approximation\cite{VCAvsSC}), which completely neglect local structural effects.
Moreover, the impurity potential is seen to affect the density of states of both compounds here analyzed, while the structural resilience of La-1111(F) prevents changes at its Fermi topology.
The comparison between the Ba-122 and the La-1111(F) compounds plays an important role in the investigation of the superconducting properties of iron pnictides.
In fact, the different behaviors found in the electronic properties of these compounds come hand in hand with different effects on the superconductivity:
Ru substitution is well known to be effective in driving Ba-122 towards superconductivity\cite{ba122expFS} while its presence is detrimental in La-1111, even in the optimally F-doped compound\cite{la1111Tc}.

\section{Conclusions}
\label{conclusions}
To summarize, we presented a detailed and systematic DFT study of Ru isovalent substitution in BaFe$_2$As$_2$ (Ba-122) and pure and F doped LaFeAsO (La-1111(F)) compounds.
The calculated internal parameters in large supercells are found to be overall in satisfying agreement with experiments, confirming the presence of magnetic order on a short range scale, which is weakened by  Ru impurities.  Effects induced by the Fe-Ru substituion also involve  the structure, the TM-As bond lengths and the TM-As-TM bond angles, though to a different extent in the Ba-122 and La-1111(F) compounds.

By disentangling the structural from the pure impurity potential effects due to Ru impurities, we found a strong dependence of the Fermi topology on the former.
A minor role is, in fact, directly played by the more delocalized Ru orbitals, that only shift the TM-As hybridization to lower energy, as clearly visible from the density of states.
Instead, the effective band structure at the Fermi level, obtained by means of the unfolding technique recently included in the VASP code, is modified accordingly to the structural changes due to Ru substitution.
In particular, the Fermi surface presents hole pockets at $\Gamma$ point shrinking as the As atoms get closer to the TM planes upon Ru substitution in Ba-122, thus resulting in an effective local electron doping compensated by further modifications at the basal $Z$-plane.
The discrepancy between DFT calculations and ARPES experiments regarding the behavior of the $d_{x^2-y^2}$ band upon Ru-substitution, appears to be related to the underestimated values of the \textit{ab initio} relaxed $z_{\rm As}$ internal parameter, that is strongly dependent on the electronic correlations.

In La-1111(F), due to its structural resilience, the Fermi topology is consistently found not to appreciably change at all.
This reflects also the well known different picture of the superconductivity scenario: 
Ru is found not to drive any superconducting state in La-1111, and even disrupts  superconductivity in the optimally F-doped compound, linearly lowering the critical temperature, while the opposite is found for the Ba-122 system.

This different behavior, suggests the need to properly take into account the local structural effects and their consequent impact on the electronic properties, in order to satisfyingly describe the peculiarities of the 122 and 1111 iron pnictide families.

\begin{acknowledgments}
This work has been carried out within a joint collaboration between the University of L'Aquila (Italy) and the University of Vienna (Austria), supported by the LLP Erasmus Placement student mobility program and the FWF-SFB ViCoM (Grant No. F41) project.
Support from the Italian Ministry for Research and Education through MIUR-PRIN2012 grant and computing time at the Vienna Scientific Cluster and ISCRA C grant at Cineca HPC are greatly acknowledged.
\end{acknowledgments}

\end{document}